\documentclass[a4paper,12pt]{article}
\usepackage[latin1]{inputenc}
\usepackage{authblk}
\usepackage{amssymb}
\usepackage{amsmath}
\usepackage{amsfonts}
\usepackage{slashed}
\usepackage{graphicx}
\usepackage{setspace}
\usepackage{fancyhdr}
\usepackage{geometry}
\usepackage{plain}
\usepackage{multirow}
\usepackage{graphicx}
\usepackage{anysize}
\usepackage{placeins}
\usepackage{float}
\usepackage{color}
\usepackage{setspace}
\usepackage[backgroundcolor=green,linecolor=green]{todonotes}
\usepackage{array}
\usepackage[colorlinks=true]{hyperref}
\usepackage[all]{hypcap}
\usepackage{dcolumn}
\usepackage{array}
 \numberwithin{equation}{section}
 \hypersetup{colorlinks, citecolor=violet, linkcolor=violet,  urlcolor=violet}
 \marginsize{2cm}{2cm}{2cm}{2cm}

\begin{document}

\thispagestyle{empty}

\begin{center}
\vspace{15mm}
\Large{\textbf{Solving Gauge Anomaly Equations in the Standard Model using the Method of Chords}} \\
\vspace{20mm}
\large\text{Dyuman Bhattacharya$^{1}$ and Sayeh Rajabi$^{2}$}\\
\vspace{15mm}

\normalsize \textit{$^1$Department of Physics and Astronomy, University of Waterloo, Waterloo, ON N2L 3G1, Canada} \\
\vspace{2mm}
\vspace{2mm}
\normalsize\textit{$^2$Department of Physics and Astronomy, University of Calgary, Calgary, AB T2N 1N4, Canada} \\

\let\thefootnote\relax\footnotetext{d7bhatta@uwaterloo.ca, s2rajabi@uwaterloo.ca}
\end{center}
\vspace{20mm}

\abstract
In a recent paper, Allanach et al. \cite{AGT1} introduced a geometric method to solve the anomaly cancellation equations for a $U(1)$ gauge theory with an arbitrary number of charges---the Method of Chords known in Diophantine analysis. We extend their result to non-Abelian gauge groups, and show that this method can be used to find the general solution to the anomaly cancellation equations for a theory with the Standard Model gauge group on a curved background. Given $K$ charges in $(\textbf{2}, \textbf{3})$, $L$ charges in $(\textbf{2}, \textbf{1})$, $M$ charges in $(\textbf{1}, \textbf{3})$, and $N$ charges in $(\textbf{1}, \textbf{1})$ representations of $SU(2)\times SU(3)$, the equations reduce to a homogeneous cubic Diophantine equation in $K+L+M+N-4$ variables.

\newpage

\section{Introduction}

Gauge anomalies must be removed for a physical theory to be mathematically consistent. This entails restricting the charges of the particles of that theory to be solutions of a set of Diophantine equations, called the anomaly cancellation equations. Diophantine equations are polynomial equations $f(x_1,x_2,\cdots,x_n)=0$ with integer coefficients and integer or rational solutions. Solving cubic Diophantine equations, which are most pertinent to particle physics, is infamously difficult, and quickly becomes intractable as the number of variables grows. \\

Costa et al. \cite{CDF1, CDF2} (CDF) have recently introduced a method to find a general solution to the anomaly cancellation equations for an arbitrary $n$ number of charges in a $U(1)$ gauge theory, given by $\sum\limits^n z_i^3 = 0$ and $\sum\limits^n z_i = 0$. Following their novel method, Allanach et al. observed a geometric interpretation in terms of cubic surfaces---known to number theorists \cite{Mordell}---for the CDF method, and reproduced their solution using the `method of chords' in projective spaces \cite{AGT1}. As the method of chords explains, given two solutions, all solutions to a homogeneous cubic Diophantine equation can be found. They further found the general solution to the anomaly cancellation equations in a $U(1)$-extension of the Standard Model gauge algebra with the Standard Model chiral fermions plus three singlets \cite{AGT2}. In another recent work \cite{DF}, a Diophantine equation in three variables was solved where the cube of the sum of variables (charges) is proportional to the sum of the cubes. These equations are crucial in finding fermion charges under $U(1)$, and hence are important in particle physics. A natural next step of these novel methods is to extend them to non-Abelian gauge groups, and in particular the Standard Model. In this note, we successfully apply the method of chords to the $G=U(1)\times SU(2)\times SU(3)$ group of the Standard Model, with the theory being on a curved background, when an arbitrary number of particles exist in the fundamental and singlet representations of the individual gauge groups. We further consider the $SU(2)\times U(1)$ electroweak theory whose matter content is $K$ left-handed chiral fermions which transform in an arbitrary $\alpha$-dimensional representation of $SU(2)$, and $L$ left-handed singlet fermions under $SU(2)$. The method of chords successfully produces all the solutions when applied to this theory. The anomaly cancellation equations in a $SU(3)\times U(1)$ theory can be similarly solved, as they are a special case of the equations in our generic $SU(2)\times U(1)$ theory where $\alpha=3$. \\

Consider a chiral gauge theory with the gauge group
\begin{equation}
\label{SMgroup}
    G=U(1)\times SU(2)\times SU(3).
\end{equation}

\noindent We shall denote the chiral matter content of the theory with $(\textbf{R}_1, \textbf{R}_2)_Y$, where $\textbf{R}_1$ and $\textbf{R}_2$ are the representations under the $SU(2)$ and $SU(3)$ groups respectively and $Y$ is the hypercharge. Furthermore, we shall consider theories with only left-handed particles without loss of generality, since we can charge conjugate any right-handed fermion, and get a left-handed one. Our theory will have $K$ charges of $(\textbf{2}, \textbf{3})_{a_i}$, $L$ charges of $(\textbf{2}, \textbf{1})_{b_i}$, $M$ charges of $(\textbf{1}, \textbf{3})_{c_i}$, and $N$ charges of $(\textbf{1}, \textbf{1})_{d_i}$. The local anomaly cancellation equations for this theory are\footnote{See \cite{Bilal,Tong,Peskin, Schwartz} for detailed discussion on the anomaly cancellation equations in the Standard Model and how to derive them.}
\begin{equation}
\label{ACE1}
    \sum_{i=1}^{K} 6a_i^3+\sum_{i=1}^{L} 2b_i^3+\sum_{i=1}^{M} 3c_i^3+\sum_{i=1}^{N} d_i^3=0,
\end{equation}
\begin{equation}
\label{ACE2}
    \sum_{i=1}^{K} 3a_i+\sum_{i=1}^{L} b_i=0,
\end{equation}
\begin{equation}
\label{ACE3}
    \sum_{i=1}^{K} 2a_i+\sum_{i=1}^{M} c_i=0,
\end{equation}
\begin{equation}
\label{ACE4}
    \sum_{i=1}^{K} 6a_i+\sum_{i=1}^{L} 2b_i+\sum_{i=1}^{M} 3c_i+\sum_{i=1}^{N} d_i=0.
\end{equation}
The first of these equations (\ref{ACE1}) comes from a triangle diagram with three external $U(1)$ gauge bosons. Hence, the anomaly cancellation condition becomes $\sum_{left} Y^3 = 0$ where we also need to multiply each term in the sum by the dimension of the representation under the non-Abelian group. The second equation (\ref{ACE2}) comes from a triangle diagram with two external $SU(2)$ gauge bosons and one external $U(1)$ gauge boson. The third equation (\ref{ACE3}) comes from a triangle diagram with two external $SU(3)$ gauge bosons and one external $U(1)$ gauge boson. In each of these two cases where we have one $U(1)$ and two $SU(N)$ gauge groups, the anomaly cancellation condition reads $\sum_{left} Y = 0$ where the only fermions that are contributing to this sum are the ones that are charged under that specific $SU(N)$ group. In order for the fermions to be coupled to gravitons, it is necessary that the mixed gauge-gravitational anomaly vanishes too. The fourth equation (\ref{ACE4}) comes from a triangle diagram with two external gravitons and one external $U(1)$ gauge boson, and is an artifact of placing the theory on curved spacetime. Note that all fermions are coupled to gravity, and therefore the anomaly cancellation condition becomes $\sum_{left} Y = 0$ with all charges contributing to the sum. The only cubic equation in this system of anomaly cancellation equations is the one that gets contribution only from the Abelian group $U(1)$, while the non-Abelian gauge groups lead to linear conditions.\\

There is one additional anomaly cancellation equation that corresponds to the triangle diagram with three external $SU(3)$ gauge bosons. In a theory with both left-handed and right-handed chiral fermions, this would add the constraint that the number of left-handed and right-handed quarks be equal. Since we have charge conjugated all right-handed particles to get a theory of only left-handed particles, this constraint becomes
\begin{equation}
\label{even2KM}
    2K+M \in \mathrm{evens}.
\end{equation}
This is equivalent to the constraint that $M \in \mathrm{evens}$. \\

We can reduce these four equations to a single cubic equation with $K+L+M+N-3$ variables. Introducing the variable
\begin{equation}
\label{z_i}
z_i=\begin{cases} 
      a_i & 1\leq i\leq K \\
      b_{i-K} & K+1\leq i\leq K+L \\
      c_{i-K-L} & K+L+1\leq i\leq K+L+M \\
      d_{i-K-L-M} & K+L+M+1\leq i\leq \delta
   \end{cases}
\end{equation}
where $\delta = K+L+M+N$ is the total number of charges, one can rewrite the anomaly cancellation equations in terms of $z_i$, and then eliminate $z_{K+L}$, $z_{K+L+M}$, and $z_{\delta}$ to get a homogeneous cubic Diophantine equation in $\delta-3$ variables,
\begin{equation}
\label{thecubic}
\begin{split}
    &\sum_{i=1}^{K} 6z_i^3+\sum_{i=K+1}^{K+L-1} 2z_i^3-2\left(\sum_{i=1}^{K} 3z_i+\sum_{i=K+1}^{K+L-1}z_i\right)^3+\sum_{i=K+L+1}^{K+L+M-1} 3z_i^3\\
    &-3\left(\sum_{i=1}^{K} 2z_i+\sum_{i=K+L+1}^{K+L+M-1}z_i\right)^3+
    \sum_{i=K+L+M+1}^{\delta-1} z_i^3+\left(\sum_{i=1}^{K} 6z_i-\sum_{i=K+L+M+1}^{\delta-1}z_i\right)^3=0.
\end{split}
\end{equation}
The $K$ sector in this equation has $K$ variables, while each other sector has $X-1$ variables, where $X=L, M, N$. Note that there are cross-terms between the $K$ sector and the other sectors in this equation, which is the consequence of our choice of eliminating $z_{K+L}$, $z_{K+L+M}$, and $z_{\delta}$. Any other choice will result in different cross terms. We will now briefly review the method of chords in the next section and apply it to equation \ref{thecubic}.

\section{Mordell's Result and the Method of Chords}

As was explained in \cite{AGT1}, in solving the \textit{homogeneous} anomaly cancellation equations, we shall use projective geometry over the field $\mathbb{Q}$ of rational numbers. For the affine space $\mathbb{Q}^n$, the projective space P${\mathbb{Q}}^{n-1}$ is ($\mathbb{Q}^n- {0})/$ $\sim$, where $\sim$ is the equivalence relation $m_1 \sim m_2$ with $m_1$, $m_2$  $\in$  $\mathbb{Q}^n$ if and only if there is some number $\lambda$  $\in$  $\mathbb{Q}$ such that $m_1 = \lambda m_2$. \\

A point in P${\mathbb{Q}}^{n-1}$ can be denoted by $[a_1,..., a_n]$ for $a_i$  $\in$  $\mathbb{Q}$. If one were to scale all the numbers in a point by the same factor, the point would remain the same. Then $d$-planes (for $d<n-1$) are defined as $d$-dimensional projective subspaces of P${\mathbb{Q}}^{n-1}$,
\begin{equation}
\label{Gamma_def}
    \Gamma = \sum_{i=1}^{d+1} \gamma_{i}p_i
\end{equation}
where $[\gamma_1:...:\gamma_{d+1}]$  $\in$ P${\mathbb{Q}}^d$ are the parameters of the plane and $p_i$  $\in$  P${\mathbb{Q}}^{n-1}$ are fixed points. We will alternatively denote planes using a generic point that lies in them. For example, a two-plane ($d=2$) that lives in P${\mathbb{Q}}^4$ ($n=5$) could be $\Gamma = [0:k:l:m:0]$, where we have five elements and three parameters. \\

Given a homogeneous polynomial equation, such as \ref{thecubic} with $n$ variables (here $\delta -3$) and rational coefficients, its rational solutions live on a hypersurface in P${\mathbb{Q}}^{n-1}$ (here P${\mathbb{Q}}^{\delta-4}$), since they can be scaled and still remain solutions. \\

It is well known that a chord between two points on a cubic hypersurface in $\mathbb{Q}^n$ will intersect the hypersurface at a third point. In other words, given two rational solutions to a cubic polynomial with rational coefficients, one can find a third. This works in the projective space P$\mathbb{Q}^n$ as well, and is called `the method of chords' \cite{AGT1,AGT2}. A result from Louis Mordell \cite{Mordell} tells us that all rational points in a cubic surface in P$\mathbb{Q}^2$ can be constructed using the method of chords, with a projective line $L$ and a point $p_1$ (that does not lie on $L$) which both lie in the surface. Any point in P$\mathbb{Q}^2$, and therefore any point on the cubic, belongs to a chord connecting $p_1$ to a point on $L$. It has been shown that Mordell's result can be extended to an arbitrary cubic hypersurface $X$ in P$\mathbb{Q}^{n-2}$ for $n\geq4$ \cite{AGT1}. The following theorem summarizes this result. \\

$\bf{Theorem}$: Starting with two disjoint planes $\Gamma_1$ and $\Gamma_2$ $\subset$ $X$ defined by a homogeneous cubic equation in the projective space $P{\mathbb{Q}}^{n}$, where $n\geq2$, every rational point $p\in$ P$\mathbb{Q}^{n}$ (and therefore every point $p\in X$) lies on a chord joining a point in $\Gamma_1$ to a point in $\Gamma_2$ with the following dimensions for the planes. If $n$ is odd,
\begin{equation}
\label{odd_n}
     d_1=d_2:=\frac{n-1}{2}
\end{equation}
where $d_1$ is the dimension of $\Gamma_1$ and $d_2$ is the dimension of $\Gamma_2$, and if $n$ is even,
\begin{equation}
\label{even_n}
\begin{split}
     d_1&:=\frac{n}{2}\\
     d_2&=d_1-1.
\end{split}
\end{equation}
We now define a projective line in $P{\mathbb{Q}}^{n}$
\begin{equation}
\label{PL}
L=\gamma_{1}p_1+\gamma_{2}p_2
\end{equation}
\noindent where $p_1 \in \Gamma_1$ and $p_2 \in \Gamma_2$ and $[\gamma_1:\gamma_2]$ $\in$ P$\mathbb{Q}^1$ are parameters that define points along the line.  An example of a 1-line in P${\mathbb{Q}}^4$ ($n=5$) is $L=k[0:1:0:0:0] + l[0:0:1:0:0]$. To find the other point where $L$ intersects $X$ (which will give us a new solution) we substitute $L$ into the cubic equation \ref{thecubic} and solve for the ratio of $\gamma_1$ and $\gamma_2$. The third point of intersection (the new solution to the cubic equation) is then constructed using points $p_1$ and $p_2$ (two known solutions of the cubic) and the ratio $[\gamma_1:\gamma_2]$ as
\begin{equation}
\label{new_z}
    z = \gamma_1 p_1 + \gamma_2 p_2.
\end{equation}
What remains is to find suitable disjoint planes $\Gamma_1$ and $\Gamma_2$ that satisfy our cubic equation. 
      
\section{Application to the Standard Model}
\label{sec:SM}
Note that upon substituting a plane of correct dimension $d$ into \ref{thecubic}, the left-hand side of the equation must \textit{trivially} vanish. This requires that the charges in the $K$ sector add up to zero, and the charges in each of the other sectors add up to zero or a single variable. If this is not the case, the left-hand side would be a non-trivial function of variables equated to zero, which would give us an additional constraint and would then reduce the dimension of the plane to $d-1$. Hence the method of chords would not be applicable. \\

There is a conflict between making the planes $\Gamma_1$ and $\Gamma_2$ simultaneously disjoint, of the right dimensions, and satisfying the cubic equation. For example, consider the case $K=L=M=N=2$. Disjoint planes of the right dimensions could be $\Gamma_1=[0:k:l:m:0]$ and $\Gamma_2=[k':-k':0:0:n']$. However, substituting the values in $\Gamma_1$ into \ref{thecubic} shows us that we must have $k=0$. This reduces the dimension of $\Gamma_1$, and renders it unsuitable for the method of chords. If instead, we had started with $\Gamma_1 = [k:-k:l:0:n]$ and $\Gamma_2 = [0:0:l':m':0]$ that satisfy \ref{thecubic} and are of the right dimensions, we would violate the requirement that the planes be disjoint. Note that it is necessary for any two corresponding sectors in different planes to be disjoint in order for the planes to be disjoint. In order for the same sectors in different planes to be disjoint, one sector must have a zero, while the other does not. In this example the $L$ sectors of $\Gamma_1$ and $\Gamma_2$ violate this requirement.  \\

One way to resolve this inconsistency, is to start with four linear conditions on the charges instead of three \ref{ACE2}--\ref{ACE4}. Then, the number of variables in the reduced cubic equation would be $\delta-4$, and the cubic hypersurface would live in $P\mathbb{Q}^{\delta-5}$. Hence, \ref{odd_n} and \ref{even_n} require that dimensions of the planes satisfy $d_1=d_2:=\frac{\delta-6}{2}$ if $\delta$ is even; and $d_1:=\frac{\delta-5}{2}$ and $d_2=d_1-1$ if $\delta$ is odd. 
\\

While any extra linear condition based on the requirements of a particular theory would suffice, we have chosen four equivalent scenarios. The three linear constraints \ref{ACE2}--\ref{ACE4} impose that if the sum of charges in a sector is zero, every other sum must vanish too. We may achieve this situation if: (a) all charges add up to zero, (b) charges of three sectors add up to zero, (c) charges of two sectors add up to zero, and (d) obviously if the sum of charges of one sector vanishes. Hence, if any of these scenarios is the case in a theory, we will get $\sum\limits^X z_i = 0$ for $X=K,L,M$ and $N$. Eliminating $z_K$, $z_{K+L}$, $z_{K+L+M}$ and $z_\delta$, the cubic equation \ref{thecubic} reduces to
\begin{equation}
\label{newcubic}
\begin{split}
    &\sum_{i=1}^{K-1} 6z_i^3 - 6\left(\sum\limits_{i=1}^{K-1}z_i\right)^3+ \sum_{i=K+1}^{K+L-1} 2z_i^3-2\left(\sum_{i=K+1}^{K+L-1}z_i\right)^3+\sum_{i=K+L+1}^{K+L+M-1} 3z_i^3\\
    &-3\left(\sum_{i=K+L+1}^{K+L+M-1}z_i\right)^3+
    \sum_{i=K+L+M+1}^{\delta-1} z_i^3-\left(\sum_{i=K+L+M+1}^{\delta-1}z_i\right)^3=0
\end{split}
\end{equation} 
with $\delta - 4$ variables which defines a cubic hypersurface in P$\mathbb{Q}^{\delta-5}$. Using \ref{newcubic}, the $[\gamma_1:\gamma_2]$ ratio \ref{PL} then becomes
\begin{equation}
\label{reduced_gamma_ratio}
\begin{split}
    [\gamma_1 : \gamma_2]=&\Biggr[\Biggr\{6\left(\sum_{i=1}^{K-1}p_{2i}^2 \ p_{1i}\right)-6\left(\sum_{i=1}^{K-1}p_{2i}\right)^2\left(\sum_{i=1}^{K-1}p_{1i}\right)\\
    &+2\left(\sum_{i=K+1}^{K+L-1}p_{2i}^2 \ p_{1i}\right)-2\left(\sum_{i=K+1}^{K+L-1}p_{2i}\right)^2\left(\sum_{i=K+1}^{K+L-1}p_{1i}\right)\\
    &+3\left(\sum_{i=K+L+1}^{K+L+M-1}p_{2i}^2 \ p_{1i}\right)-3\left(\sum_{i=K+L+1}^{K+L+M-1}p_{2i}\right)^2\left(\sum_{i=K+L+1}^{K+L+M-1}p_{1i}\right)\\
    &+\sum_{i=K+L+M+1}^{\delta-1}p_{2i}^2 \ p_{1i}-\left(\sum_{i=K+L+M+1}^{\delta-1}p_{2i}\right)^2\left(\sum_{i=K+L+M+1}^{\delta-1}p_{1i}\right)\Biggr\}\\
    &: -\Biggr\{ p_1 \leftrightarrow p_2 \Biggr\}\Biggr]
\end{split}
\end{equation}
where in the last line $p_a$'s are swapped. \\

Given the structure of \ref{newcubic}, to find $\Gamma_1$ and $\Gamma_2$, we may consider each $K, L, M$ and $N$ sector independently, noting that the planes must be of the right dimensions. Furthermore, one must make sure to order the elements in the planes in such a way that \ref{new_z} does not only produce vector-like solutions. Although the method of chords produces all solutions, we are mostly interested in non-vector-like solutions---those that satisfy $z_i+z_j \neq 0$ for all $i$ and $j$ in a sector. Note that there are $X-1$ elements in each $X=K,L,M,N$ sector. For sector $X$, depending on whether $X$ is even or odd, possible sets of $X-1$ ordered elements that can make the sector disjoint are
\begin{equation}
\label{sectors}
\begin{split}
    \Gamma^{e1}&=[0:x_1:\cdots:x_{\frac{X}{2}-1}:-x_1:\cdots:-x_{\frac{X}{2}-1}] \\
    \Gamma^{e2}&=[x'_1:\cdots:x'_{\frac{X}{2}}:-x'_1:\cdots:-x'_{\frac{X}{2}-1}] \\
    \Gamma^{o1}&=[0:x_1:\cdots:x_{\frac{X-3}{2}}:-x_1:\cdots:-x_{\frac{X-1}{2}}] \\
    \Gamma^{o2}&=[x'_1:\cdots:x'_{\frac{X-1}{2}}:-x'_1:\cdots:-x'_{\frac{X-1}{2}}].
\end{split}
\end{equation}
Superscripts $e$ and $o$ indicate even and odd values of $X$ respectively. The planes are now constructed by merging these sectors in such a way that their dimensions satisfy \ref{odd_n} and \ref{even_n}. Since $M$ is always even, there are in total eight possible combinations of even and odd parameters ($K,L,M,N$) for which we present example planes as follows. Without loss of generality, we can assume that the parameters are all non-zero. If one sector is empty, one has to return to the set of anomaly cancellation equations, construct a new reduced cubic and accordingly make new planes. For the case of two parameters being zero, the problem reduces to solving the anomaly equations in a $SU(2)\times U(1)$ theory, explained in section \ref{sec:electroweak}. For the case of three parameters being zero, the problem reduces to solving the anomaly equations in a $U(1)$ gauge theory \cite{AGT1}.\\

For the case of all even $K, L, M, N$, we pick up our $K$ sector from the first type of $\Gamma_1$, namely $\Gamma^{e1}$, substituting $K$ for $X$. The $L$ and $M$ sectors follow the pattern in $\Gamma^{e2}$, and for the $N$ sector we choose the pattern of $\Gamma^{e1}$ to make the dimension of the plane $\frac{\delta - 6}{2}$. To construct $\Gamma_2$, one must ensure that the two planes are disjoint. This entails exchanging each $\Gamma^{e1}$ with $\Gamma^{e2}$, so that corresponding sectors have no overlap other than at the origin. We summarize our prescriptions as follows, omitting the subscripts $K, L, M, N$ for brevity. Note that the addition sign here signifies merging the sectors.
\begin{equation}
\label{alleven}
\begin{split}
    \Gamma_1 &= \Gamma^{e1} + \Gamma^{e2} + \Gamma^{e2} + \Gamma^{e1} \\
    \Gamma_2 &= \Gamma^{e2} + \Gamma^{e1} + \Gamma^{e1} + \Gamma^{e2}.
\end{split}
\end{equation}
We construct the rest of the planes similarly. For the case of even $K, L, M$ and odd $N$,
\begin{equation}
\label{oddN}
\begin{split}
    \Gamma_1 &= \Gamma^{e1} + \Gamma^{e2} + \Gamma^{e2} + \Gamma^{o1}\\
    \Gamma_2 &= \Gamma^{e2} + \Gamma^{e1} + \Gamma^{e1} + \Gamma^{o2}.
\end{split}
\end{equation}
For the case of even $K, M, N$ and odd $L$,
\begin{equation}
\label{oddL}
\begin{split}
    \Gamma_1 &= \Gamma^{e1} + \Gamma^{o1} + \Gamma^{e2} + \Gamma^{e2} \\
    \Gamma_2 &= \Gamma^{e2} + \Gamma^{o2} + \Gamma^{e1} + \Gamma^{e1}.
\end{split}
\end{equation}
For the case of even $K, M$ and odd $L, N$,
\begin{equation}
\label{oddLN}
\begin{split}
    \Gamma_1 &= \Gamma^{e1} + \Gamma^{o1} + \Gamma^{e2} + \Gamma^{o1} \\
    \Gamma_2 &= \Gamma^{e2} + \Gamma^{o2} + \Gamma^{e1} + \Gamma^{o2}.
\end{split}
\end{equation}
For the case of even $L, M, N$ and odd $K$,
\begin{equation}
\label{oddK}
\begin{split}
    \Gamma_1 &= \Gamma^{o1} + \Gamma^{e1} + \Gamma^{e2} + \Gamma^{e2} \\
    \Gamma_2 &= \Gamma^{o2} + \Gamma^{e2} + \Gamma^{e1} + \Gamma^{e1}.
\end{split}
\end{equation}
For the case of even $L, M$ and odd $K, N$,
\begin{equation}
\label{oddKN}
\begin{split}
    \Gamma_1 &= \Gamma^{o1} + \Gamma^{e1} + \Gamma^{e2} + \Gamma^{o2} \\
    \Gamma_2 &= \Gamma^{o2} + \Gamma^{e2} + \Gamma^{e1} + \Gamma^{o1}.
\end{split}
\end{equation}
For the case of even $M, N$ and odd $K, L$,
\begin{equation}
\label{oddKL}
\begin{split}
    \Gamma_1 &= \Gamma^{o1} + \Gamma^{o2} + \Gamma^{e2} + \Gamma^{e1} \\
    \Gamma_2 &= \Gamma^{o2} + \Gamma^{o1} + \Gamma^{e1} + \Gamma^{e2}.
\end{split}
\end{equation}
For the case of even $M$ and odd $K,L,N$,
\begin{equation}
\label{oddKLN}
\begin{split}
    \Gamma_1 &= \Gamma^{o1} + \Gamma^{o2} + \Gamma^{e2} + \Gamma^{o1} \\
    \Gamma_2 &= \Gamma^{o2} + \Gamma^{o1} + \Gamma^{e1} + \Gamma^{o2}.
\end{split}
\end{equation}
To better illustrate these solutions, consider equation \ref{newcubic} with even $L$ and $M$, and odd $K$ and $N$. Planes can be constructed as prescribed in \ref{oddKN},
\begin{equation}
\begin{split}
    \Gamma_1 = &[0:k_1:\cdots:k_{\frac{K-3}{2}}:-k_1:\cdots:-k_{\frac{K-1}{2}}:0:l_1:\cdots:l_{\frac{L}{2}-1}:-l_1:\cdots:-l_{\frac{L}{2}-1} \\ 
    &:m'_1:\cdots:m'_{\frac{M}{2}}:-m'_1:\cdots:-m'_{\frac{M}{2}-1}:n'_1:\cdots:n'_{\frac{N-1}{2}}:-n'_1:\cdots:-n'_{\frac{N-1}{2}}]\\
    \Gamma_2=
    &[k'_1:\cdots:k'_{\frac{K-1}{2}}:-k'_1:\cdots:-k'_{\frac{K-1}{2}}:
    l'_1:\cdots:l'_{\frac{L}{2}}:-l'_1:\cdots:-l'_{\frac{L}{2}-1} \\
    &:0:m_1:\cdots:m_{\frac{M}{2}-1}:-m_1:\cdots:-m_{\frac{M}{2}-1}:
    0:n_1:\cdots:n_{\frac{N-3}{2}}:-n_1:\cdots:-n_{\frac{N-1}{2}}].
\end{split}
\end{equation}
For example if $K=5, L=4, M=6$ and $N=3$, choosing the following points from $\Gamma_1$ and $\Gamma_2$,
\begin{equation}
\begin{split}
    p_1 &= [0: 1: -1: 1: 0: 1: -1: 1: 2: -2: -1: -2: -3: 3] \\
    p_2 &= [-2: 0: 2: 0: 1: 0: -1: 0: -3: 0: 3: 0: 0: -1]
\end{split}
\end{equation}
one can find the third point of intersection (the new solution) at
\begin{equation}
    z = [-26: 2: 24: 2: -2: 13: 2: -15: 0: 2: -35: -4: 37: -4: 4: -6: -7: 13],
\end{equation}
which includes the eliminated variables $z_5, z_9, z_{15}$ and $z_{18}$ too.\\

The method of chords will only work for $\delta\geq 7$. For smaller $\delta$, we examine the solutions to \ref{ACE1}--\ref{ACE4} together with the fourth condition, as follows. For cases with $\delta = 6, 5$ or $4$, if only one sector is non-empty, e.g. $(K,L,M,N) = (0,\delta,0,0)$, the problem reduces to the $U(1)$ anomaly cancellation solved in \cite{AGT1}. For all values of $\delta$, if a sector has a single variable, the four linear conditions require the single variable to be zero. For example, $(K,L,M,N)=(4,1,0,1)$ can be solved as explained in \cite{AGT1} for the variables in the $K$ sector, while the other two variables $z_5$ and $z_6$ vanish.  If there are two variables in a sector, linear constraints require that they be vector-like $[1:-1]$. For each sector that contains three variables the solutions will be $[1:-1:0]$, $[1:0:-1]$ and $[0:1:-1]$ as similarly reported in \cite{AGT1}. These observations help us immediately find all solutions for all possible permutations of the parameters (values of $K$, $L$, $M$ and $N$). For example, when $\delta=6$ the solutions for $(3,0,2,1)$ will be found by merging the solutions of the $K$, $M$ and $N$ sectors: $[k:-k:0:m:-m:0]$, $[k:0:-k:m:-m:0]$ and $[0:k:-k:m:-m:0]$ where $k, m \in \mathbb{Q}$.

\section{Application to $SU(2)\times U(1)$}
\label{sec:electroweak}
The local anomaly equations for a $SU(2)\times U(1)$ gauge theory on a curved background with $K$ left-handed chiral particles with charges $x_i$ that transform in the $\alpha$-dimensional representation of $SU(2)$, and $L$ left-handed chiral particles that transform in the singlet representation of $SU(2)$ are
\begin{equation}
\label{ACE2-1}
\sum_{i=1}^{K} \alpha{x_i}^3+\sum_{i=1}^{L}{y_i}^3=0,
\end{equation}
\begin{equation}
\label{ACE2-2}
\sum_{i=1}^{K} x_i=0,
\end{equation}
\begin{equation}
\label{ACE2-3}
\sum_{i=1}^{K} \alpha{x_i}+\sum_{i=1}^{L}{y_i}=0.
\end{equation}
Note that for the special case of $K=1$, or $L=1$, the single element is equal to zero, and the problem reduces to the case of $U(1)$ gauge anomaly cancellation \cite{AGT1}.\\

The system of equations \ref{ACE2-1}--\ref{ACE2-3} can be reduced to the single equation
\begin{equation}
\label{ACE2-Single}
\alpha\left[\sum_{i=1}^{K-1} {z_i}^3-\left(\sum_{i=1}^{K-1} {z_i}\right)^3\right]+\sum_{i=K}^{K+L-2} {z_i}^3-\left(\sum_{i=K}^{K+L-2} {z_i}\right)^3=0.
\end{equation}
As a cubic homogeneous polynomial, its rational solutions can be found using the method of chords in $P{\mathbb{Q}}^{K+L-3}$. If $K+L$ is even, then $d_1=d_2=\frac{K+L-4}{2}$, where $d_1$ and $d_2$ are the dimensions of the planes $\Gamma_1$ and $\Gamma_2$ respectively. If $K+L$ is odd, then $d_1=\frac{K+L-3}{2}$ and $d_2=d_1-1$. Similar calculations to those in the previous section yield the ratio
\begin{equation}
\label{GR}
[\gamma_1:\gamma_2]=\left[\sum_{i=1}^{K+L-2} p_{1i} \ T_{2i}: -\sum_{i=1}^{K+L-2}p_{2i} \ T_{1i}\right]
\end{equation}
where
\begin{equation}
T_{ai}=\begin{cases} 
      \alpha\left[p_{ai}^2-\left(\sum\limits_{j=1}^{K-1}p_{aj}\right)^2\right] & 1\leq i\leq K-1 \\
      p_{ai}^2-\left(\sum\limits_{j=K}^{K+L-2}p_{aj}\right)^2 & K\leq i\leq K+L-2. 
   \end{cases}
\end{equation}

\noindent For cases with even $K$ and $L$, suitable planes are
\begin{equation}
\begin{split}
     \Gamma_1^{ee}&=[k_1:\cdots:k_{\frac{K}{2}}:-k_1:\cdots:-k_{\frac{K}{2}-1}:0:l_1:\cdots:l_{\frac{L}{2}-1}:-l_1:\cdots:-l_{\frac{L}{2}-1}]\\
     \Gamma_2^{ee}&=[0:k'_1:\cdots:k'_{\frac{K}{2}-1}:-k'_1:\cdots:-k_{\frac{K}{2}-1}:l'_1:\cdots:l'_{\frac{L}{2}}:-l'_1:\cdots:-l'_{\frac{L}{2}-1}].
\end{split}
\end{equation}
Suitable planes for cases with even $K$ and odd $L$ are
\begin{equation}
\begin{split}
     \Gamma_1^{eo}&=[k_1:\cdots:k_{\frac{K}{2}}:-k_1:\cdots:-k_{\frac{K}{2}-1}:l_1:\cdots:l_{\frac{L-1}{2}}:-l_1:\cdots:-l_{\frac{L-1}{2}}] \\
     \Gamma_2^{eo}&=[0:k'_1:\cdots:k'_{\frac{K}{2}-1}:-k'_1:\cdots:-k'_{\frac{K}{2}-1}:0:l'_1:\cdots:l'_{\frac{L-3}{2}}:-l'_1:\cdots:-l'_{\frac{L-1}{2}}].
\end{split}
\end{equation}
For cases with odd $K$ and even $L$,
\begin{equation}
\begin{split}
     \Gamma_1^{oe} &= [k_1:\cdots:k_{\frac{K-1}{2}}:-k_1:\cdots:-k_{\frac{K-1}{2}}:l_1:\cdots:l_{\frac{L}{2}}:-l_1:\cdots:-l_{\frac{L}{2}-1}] \\
     \Gamma_2^{oe} &= [0:k'_1:\cdots:k'_{\frac{K-3}{2}}:-k'_1:\cdots:-k'_{\frac{K-1}{2}}:0:l'_1:\cdots:l'_{\frac{L}{2}-1}:-l'_1:\cdots:-l'_{\frac{L}{2}-1}].
\end{split}
\end{equation}
For cases with odd $K$ and $L$,
\begin{equation}
\begin{split}
     \Gamma_1^{oo} &= [k_1:\cdots:k_{\frac{K-1}{2}}:-k_1:\cdots:-k_{\frac{K-1}{2}}:l_1:\cdots:l_{\frac{L-1}{2}}:-l_1:\cdots:-l_{\frac{L-1}{2}}] \\
     \Gamma_2^{oo} &= [0:k'_1:\cdots:k'_{\frac{K-3}{2}}:-k'_1:\cdots:-k'_{\frac{K-1}{2}}:0:l'_1:\cdots:l'_{\frac{L-3}{2}}:-l'_1:\cdots:-l'_{\frac{L-1}{2}}].
\end{split}
\end{equation}
These planes are all disjoint, and have the right dimensions, so by the theorem, they allow for all rational solutions to the cubic equation \ref{ACE2-Single} to be found. \\

The method of chords can only work for $K+L\geq5$. For smaller $K+L$, solutions can be easily found similar to the special cases studied in section \ref{sec:SM}. If a sector contains one variable, it must be zero. Two variables in a sector give only vector-like solutions, and with three variables, one gets the three solutions: $[x_1:x_2:x_3]=[1:0:-1], [0:1:-1],$ and $[1:-1:0]$. Finally, when $(K,L)=(4,0)$ or $(0,4)$, the problem reduces to the case of $U(1)$ gauge anomaly cancellation \cite{AGT1}. Merging the solutions of the two sectors yields the full solution. For example, with $(K,L)=(2,2)$, the solutions will be $[x_1:x_2:y_1:y_2]=[k:-k:l:-l]$ where $k$ and $l\in\mathbb{Q}$. \\

The anomaly cancellation equations can also apply for an $SU(3)\times U(1)$ gauge theory that describes strong and electromagnetic interactions. This can be done by setting $\alpha=3$ in \ref{ACE2-1} and \ref{ACE2-3}. Now, the theory has $K$ particles that transform in the fundamental representation of $SU(3)$, and $L$ particles that transform as singlets under $SU(3)$. The only additional constraint comes from the Feynman diagram with three external $SU(3)$ bosons. This adds the condition that  that the theory has an equal number of left-handed and right-handed fermions that transform in the same representation of $SU(3)$. In our general theory, which has $K$ left-handed particles that do not transform as singlets, this means that $K$ must be even (since right-handed particles become left-handed particles after charge conjugation).

\section{Concluding Remarks}

We showed that the method of chords in projective space can be used to find the rational solutions to the anomaly equations for a general Standard Model gauge theory with particles that transform in the singlet and fundamental representations of $SU(2)$ and $SU(3)$. In order to find suitable planes, we imposed an extra constraint on charges that together with \ref{ACE2}--\ref{ACE4} gave us four linear conditions. As the charges in this problem were grouped into four sectors, the four conditions could help us treat each sector in the planes individually---noting that the planes must be of right dimension and produce non-vector-like solutions when they exist. The need for an additional condition is more evident when this set of anomaly equations is compared to the ones in other theories solved by the method of chords. The $U(1)$ anomaly cancellation problem \cite{AGT1} had one sector and one linear constraint. Similarly, there were two linear conditions for the case of $SU(2)\times U(1)$ (and $SU(3)\times U(1)$) in which there were two sectors of charges. Our Standard Model gauge theory has four sectors of charges, but only three linear conditions. In order to satisfy the requirements that planes be disjoint, of the right dimensions, and be inside the cubic hypersurface, it is necessary to add a fourth linear condition.\\

In addition, the method of chords was used to find the solutions to the anomaly equations for a generic $SU(2) \times U(1)$ gauge theory describing electroweak interactions. This success opens the door to further advancements in using methods from number theory to solve problems in theoretical physics. The next step would be to further generalize the theory, by having each particle transform in its own individual representations under $SU(2)$ and $SU(3)$. As the resulting anomaly equation would be a cubic, its solutions could be found using the method of chords. These models are interesting to study as they could be candidates in physics beyond the Standard Model. \\

Other extensions of our work could be to probe the space of solutions of the cubic equation \ref{newcubic} and to analyze the solutions in particular limits of the parameters. In particular, we are interested in learning the `large-$N$' limit of the Diophantine equations describing the elimination of gauge anomalies from the theory. Statistical analysis of the solutions could be a means to reveal more connections between number theory and physics.

\section*{Acknowledgments}
DB is grateful to Ben Gripaios for motivation, and numerous discussions at the earliest stages of this work.

\end{document}